\documentclass[aps,prd]{revtex4}
\pdfoutput=1
\usepackage{subfigure,graphics} 
\usepackage{flafter} 
\begin{document} 
\title{
Non-gaussianity in the strong regime of warm inflation} 
\author{Ian G. Moss}
\email{ian.moss@ncl.ac.uk}
\author{Timothy Yeomans}
\email{timothy.yeomans@ncl.ac.uk}
\affiliation{School of Mathematics and Statistics, University of  
Newcastle Upon Tyne, NE1 7RU, UK}

\date{\today}


\begin{abstract}
The bispectrum of scalar mode density perturbations is analysed for the
strong regime of warm inflationary models. This analysis generalises previous results
by allowing damping terms in the inflaton equation of motion that are dependent on temperature. A
significant amount of non-gaussianity emerges with constant (or local) non-linearity
parameter $f_{NL}\sim 20$, in addition to the terms with non-constant $f_{NL}$ which are
characteristic of warm inflation.

\end{abstract}
\pacs{PACS number(s): }

\maketitle
\section{introduction}

Observations of the cosmic microwave background are consistent with the
existence of gaussian, weakly scale dependent, density perturbations as
predicted by most inflationary 
models \cite{Mukhanov:1981xt,guth82,hawking82-2,bardeen83}. 
The amount of non-gaussianity produced
by the simplest inflationary models is small and unlikely to be to be
observable by the next generation of experiments, but this still leaves open
the possibility that a slightly more exotic inflationary model could produce a
measurable effect.

One variation on inflation is the warm inflationary scenario \cite{berera95}
(see also \cite{moss85} and the review \cite{Berera:2008ar}). The non-gaussianity 
produced by a warm inflationary
scenario with constant friction coefficient has been presented elsewhere
\cite{Moss:2007cv}. The bispectrum of the non-gaussianity is large,
and it has a distinctive dependence on wave number. In this paper, we shall examine
the bispectrum of the non-gaussianity in the more likely situation where
the friction coefficient is not constant.
We shall see that the bispectrum has two terms, one which is like the previous
warm inflation bispectrum and a new term which is typical of density perturbations 
with local non-linearities.

Warm inflation is characterised by the rate of radiation production during the 
inflationary era. The radiation can affect both the homogeneous evolution of the 
inflaton field and the inhomogeneous fluctuations. If the radiation field has a 
strong damping effect on the inflaton dynamics, then we have what is known as 
the strong regime of warm inflation. If the damping effect is small, but the fluctuations are still
influenced by radiation, then we
have the weak regime of warm inflation. In both strong and weak regimes of warm inflation, 
fluctuations in the radiation are transfered to the inflaton   
\cite{berera96,Lee:1999iv,berera97,berera98,berera00,taylor00,DeOliveira:2001he,Hall:2003zp} 
and become the primary source of density fluctuations. This is the most
significant difference between warm inflation and traditional cold inflation.

The simplest warm inflationary scenarios assume that the radiation produced
during the inflationary era thermalises at a rate faster than the expansion rate. 
This type of warm inflationary model is therefore rather restrictive. However, the possibility of
thermalisation occurring is enhanced by inflaton decay channels, which are naturally present in
many supersymmetric theories, where the inflaton decays into light radiation fields
through heavy particle intermediaries 
\cite{berera02,Berera:2004kc,Bastero-Gil:2005gy,BasteroGil:2009ec}. 
In these models, the damping of the inflaton field is described by a friction
coefficient $\Gamma\propto T^c$, where  $T$ is the 
temperature. In particular, $c\approx 3$ at temperatures small compared to the heavy particle masses
\cite{Moss:2006gt,BasteroGil:2010pb}. 
The temperature dependence has been found to have a large effect on the size of the density
fluctuations \cite{Graham:2009bf}, and
we shall examine now how the temperature dependence affects the non-gaussianity.

A feature of warm inflation is that a significant amount of non-gaussianity is produced whilst the
density fluctuations are
still on sub-horizon scales. In the previous analysis, this was due to the non-linearity caused by
the bulk velocity of the radiation. In the new analysis, there are terms in the non-gaussianity
which are proportional to the parameter $c$. The only restriction on $c$ for the consistency of warm
inflation is
that $c<4$ \cite{Moss:2008yb}. This leads to effects which are large compared to slow-roll
parameters. For comparison, the non-linearities produced by derivatives of the
inflaton potential in warm inflationary models was looked at by Gupta et al.
\cite{Gupta:2002kn,Gupta:2005nh}, but their contribution to the non-gaussianity is governed by
slow-roll parameters and it is tiny in comparison with the true non-gaussianity produced in warm
inflation.

Fluctuations in the cosmic microwave background provide an observational link to
the density fluctuations at the surface of last scattering. We know, in principle, how to
evolve these fluctuations from early times using, for example, the Bardeen
variable $\zeta$ \cite{Bardeen:1980kt}. Observations can be compared to
predictions for various
moments of the probability distribution of $\zeta$. The most important of
these is the primordial power spectrum of fluctuations $P_\zeta(k)$, defined
by the stochastic average
\begin{equation}
\langle \zeta({\bf k}_1)\zeta({\bf k}_2)\rangle=(2\pi)^3P_\zeta(k_1)
\delta^3({\bf k}_1+{\bf k}_2).\label{defps}
\end{equation}
The bispectrum $B_\zeta(k_1,k_2,k_3)$, defined by 
\begin{equation}
\langle \zeta({\bf k}_1)\zeta({\bf k}_2)\zeta({\bf k}_3)\rangle=
(2\pi)^3B_\zeta(k_1,k_2,k_3)
\delta^3({\bf k}_1+{\bf k}_2+{\bf k}_3),
\end{equation}
can be used to examine the non-gaussianity in the density fluctuations. 
The normalised amount of non-gaussianity in the bispectrum is
described by a non-linearity function $f_{NL}$, defined by
\begin{equation}
f_{NL}(k_1,k_2,k_3)={5\over 6}{B_\zeta(k_1,k_2,k_3)\over
P_\zeta(k_1)P_\zeta(k_2)+P_\zeta(k_2)P\zeta(k_3)+P_\zeta(k_3)P_\zeta(k_1)}.
\label{fdef}
\end{equation}
where the $5/6$ factor is convenient for cosmic microwave background
comparisons \cite{Komatsu:2001rj}. Models with constant $f_{NL}$ are often 
called local models because this type of non-gaussianity can arise from local 
non-linearities in the density perturbations.

Non-linear evolution during the inflationary era can result in non-gaussianity appearing in
the primordial density fluctuations. The amount of non-gaussianity produced by
vacuum fluctuations in single-field inflationary models is typically around a few per cent
\cite{Falk:1992sf,Gangui:1993tt,Acquaviva:2002ud},
and can be related to the standard set of inflationary slow-roll parameters \cite{liddle}. 
This is small compared to non-inflationary effects. For
example, the second order Sachs-Wolfe effect is expected to act as a source
of non-gaussianity in the cosmic microwave background observations equivalent
to $f_{NL}\sim 1$ \cite{Munshi:1995eh,Pyne:1995bs}. 

Models of inflation with multiple scalar fields, acting as sources of density 
fluctuations in the curvaton scenario \cite{Linde:1996gt,Lyth:2003ip}, 
or modifying the reheating phase of the universe \cite{Zaldarriaga:2003my}, 
can produce a level of non-gaussianity above the foreground effects, even significantly above
the foreground for particular parameter choices.  Non-gaussianity can also be produced by
modifications 
to the kinetic part of the inflaton Lagrangian in D-brane models 
\cite{Silverstein:2003hf,Alishahiha:2004eh}. This type of non-gaussianity is concentrated on
equilateral wave-vector triangles $k_1,k_2,k_3$, unlike the local form, which is concentrated
on oblique triangles.

The best observational limit on the non-gaussianity at present is from
the WMAP seven-year data release \cite{Komatsu:2010fb}, which gives $-10<f^{local}_{NL}<74$ with 
$95\%$ confidence for a constant (or local) component.
The Planck satellite observations may have a sensitivity limit of around
$|f^{local}_{NL}|\sim5-10$, depending on how well the signal can be separated from the galactic
foreground
\cite{Komatsu:2001rj}. Our prediction for $f^{local}_{NL}$ from the strong regime of warm
inflation lies well above the Planck detection threshold in most models, but the most
significant feature is the presence of a term in the bispectrum which could eventually provide a
means to distinguish warm inflation from other sources of non-gaussianity. The prospects
for observing this term have been discussed in Refs. \cite{Moss:2007qd} and \cite{Fergusson:2008ra}.

The paper is organised as follows. We begin in section II with a brief
introduction to the notion of warm inflation. In section III, we introduce
fluctuations of the inflaton field described by a Langevin equation and expand 
the fluctuations to second order. The limit of strong dissipation is introduced in 
section IV, and the bispectrum is calculated in section V. Some observational prospects
are discussed in the conclusion. We use units with $\hbar=c=1$.

\section{warm inflation}

Warm inflation occurs when there is a significant amount of particle production
during the inflationary era. We shall assume that
the particle interactions are strong enough to produce a thermal gas of
radiation with temperature $T$. In this case, warm inflation will occur when $T$ is
larger than the energy scale set by the expansion rate $H$. The production of
radiation is associated with a damping effect on the inflaton, whose equation
of motion becomes
\begin{equation}
\ddot\phi+(3H+\Gamma)\dot\phi+V_\phi=0
\end{equation}
where $\Gamma(\phi,T)$ is a friction coefficient, $H$ is the Hubble parameter
and $V_\phi$ is the $\phi$ derivative of the inflaton potential $V(\phi,T)$.

The effectiveness of warm inflation can be parameterised by a parameter $r$,
defined by
\begin{equation}
r={\Gamma\over 3H}
\end{equation}
When $r\gg 1$ the warm inflation is described as being in the strong regime and when $r\ll1$ the
warm
inflation is in the weak regime.

Consistent models of warm inflation \cite{Moss:2008yb} require a suppression of thermal corrections
to the inflaton potential, so that the effective potential separates into inflaton and radiation
components
\begin{equation}
V(\phi,T)=V(\phi)+\rho_r(T),
\end{equation}
where $\rho_r$ is the radiation density
\begin{equation}
\rho_r={\pi^2\over 30}g_*T^4.
\end{equation}  
In this case, the time evolution is described by the equations
\begin{eqnarray}
&&\ddot\phi+(3H+\Gamma)\dot\phi+V_\phi=0,\\
&&\dot\rho_r+4H\rho_r=\Gamma\dot\phi^2,\\
&&3H^2=4\pi G\left(2V+2\rho_r+\dot\phi^2\right)
\end{eqnarray}
During inflation we apply a slow-roll approximation and drop the highest
derivative terms in the equations of motion,
\begin{eqnarray}
&&3H(1+r)\dot\phi+V_\phi=0,\label{sr1}\\
&&4H\rho_r=\Gamma\dot\phi^2,\label{sr2}\\
&&3H^2=8\pi G V\label{sr3}
\end{eqnarray}
The validity of the slow-roll approximation depends on the slow-roll parameters
defined in \cite{Hall:2003zp},
\begin{equation}
\epsilon={1\over 16\pi G}\left({V_\phi\over V}\right)^2,\qquad
\eta={1\over 8\pi G}\left({V_{\phi\phi}\over V}\right),\qquad
\beta={1\over 8\pi G}\left({\Gamma_\phi V_\phi\over \Gamma V}\right)
\label{slowrp}
\end{equation}
The slow-roll approximation holds when $\epsilon\ll 1+r$, $\eta\ll 1+r$ and
$\beta\ll 1+r$. Any quantity of order $\epsilon/(1+r)$ will be described as being
first order in the slow-roll approximation.

The temperature dependence of the friction coefficient $\Gamma$ plays an important role
in the present analysis. We parameterise this by a parameter $c$,
\begin{equation}
c={T\,\Gamma_T\over \Gamma},
\end{equation}
where $\Gamma_T$ denotes the $T$ derivative of $\Gamma$. This parameter is not necessarily small,
but a stability analysis of warm inflation shows that $c<4$ for a consistent model
\cite{Moss:2008yb}.

\section{Fluctuations}

Thermal fluctuations are the main source of density perturbations in warm
inflation. Thermal noise is transfered to the inflaton field; mostly on small
scales. As the comoving wavelength of a perturbation expands, the thermal
effects decrease until the fluctuation amplitude freezes out \cite{berera00}. 
In the strong regime of warm inflation, this occurs
when the wavelength of the fluctuation is still small in comparison with
the size of the cosmological horizon.

\subsection{Inflaton fluctuations}

The behaviour of a scalar field interacting with radiation can be analysed
using the Schwinger-Keldysh approach to non-equilibrium field theory 
\cite{schwinger61,keldysh64}. In flat spacetime, when the small-scale behaviour of the 
fields is averaged out, a simple picture emerges in which the field can
be described by a stochastic system whose evolution is determined by a Langevin
equation \cite{calzetta88}. This takes the form
\begin{equation}
-\nabla^2\phi(x,t)+\Gamma\dot\phi(x,t)+V_\phi=(2\Gamma T)^{1/2}\xi(x,t),\label{stoc}
\end{equation}
where $\nabla^2$ is the flat spacetime Laplacian and $\xi$ is a stochastic
source. For a weakly interacting radiation gas the probability distribution of the source term can
be approximated by a localised gaussian distribution with correlation function 
\cite{gleiser94,Berera:2007qm},
\begin{equation}
\langle\xi(x,t)\xi(x',t')\rangle=\delta^{(3)}(x-x')\delta(t-t').
\end{equation}
We shall restrict ourselves to this gaussian noise approximation.

We can use the equivalence principle to adapt the flat spacetime Langevin
equation to an expanding universe during a period of warm inflation by replacing ordinary 
derivatives with covariant derivatives in the cosmological metric
with scale factor $a$ and co-moving coordinates $x^\alpha$. The Langevin
equation will retain its local form as long as the microphysical and thermal scales in the problem
are small compared to the cosmological scale \cite{Berera:2007qm,Berera:2008ar}. However, the rest
frame of the fluid will have a non-zero $3-$velocity with respect to the
cosmological frame and we must include an advection term. Another alteration is that
the coefficient of the noise term is changed slightly by the expansion 
(for details see ref \cite{Graham:2009bf}),
to $K=(2\Gamma_{\rm eff} T)^{1/2}$, where $\Gamma_{\rm eff}=\Gamma+H$.

The Langevin equation becomes \cite{Moss:2007cv}
\begin{equation}
\ddot\phi(x,t)+3H\dot\phi(x,t)+\Gamma D\phi
+V_\phi-a^{-2}\partial^2\phi(x,t)=K\xi(x,t)\label{langa}
\end{equation}
where $\partial^2$ is the Laplacian in the expanding frame and $D\phi$ is the derivative
along the radiation fluid. The correlation function for the noise, expressed in terms of the 
comoving cosmological coordinates, has the form
\begin{equation}
\langle\xi(x,t)\xi(x',t')\rangle= a^{-3}(2\pi)^2
\delta^{(3)}(x-x')\delta(t-t').
\end{equation}
The fluid velocity components in a coordinate frame are $u=(\gamma,u^\alpha)$, where
$\gamma$ is the Lorentz factor. With this choice of components,
\begin{equation}
D\phi=\gamma\dot\phi+u^\alpha\partial_\alpha\phi.
\end{equation}

The inflaton will generate metric inhomogeneities, but with a suitable choice of
gauge, these can be discarded on sub-horizon scales (see Sect. \ref{metricsec}). 
We shall use a uniform expansion rate gauge. Eq. (\ref{langa}) applies on
scales which are intermediate between the thermal averaging scale and the horizon scale.
Later, we use a matching argument to extend the fluctuations to large scales.

The analysis of the Langevin
equation can be simplified by introducing a new
time coordinate $\tau=(aH)^{-1}$ and using the slow-roll approximation. 
We are
led to the equation
\begin{equation}
\phi^{\prime\prime}(x,\tau)-(3\gamma r+2)\tau^{-1}\phi'(x,\tau)-
3r\tau^{-1}a\,u^\alpha\partial_\alpha\phi(x,\tau)-\partial^2\phi(x,\tau)
=K\hat\xi(x,\tau)\label{langb}
\end{equation}
where a prime denotes a derivative with respect to $\tau$ and we have kept only
the leading terms in the slow-roll approximation. The noise term has been
rescaled so that its correlation function is now
\begin{equation}
\langle\hat\xi(x,\tau)\hat\xi(x',\tau')\rangle=(2\pi)^3
\delta^{(3)}(x-x')\delta(\tau-\tau').\label{cf}
\end{equation}
This equation is non-linear because $\Gamma$, $u^\alpha$ and $K$ depend on
$\phi$ and $T$.

Now we treat the source term as a small perturbation and expand the inflaton
field
\begin{equation}
\phi(x,\tau)=\phi(\tau)+\delta_1\phi(x,\tau)+\delta_2\phi(x,\tau)+\dots
\label{ipe}
\end{equation}
where $\delta_1\phi$ is the linear response due to the source $\hat\xi$.
Similarly, for the fluid velocity,
\begin{equation}
u^\alpha(x,\tau)=u_1^\alpha(x,\tau)+u_2^\alpha(x,\tau)+\dots
\label{upe}
\end{equation}
This expansion is substituted into the langevin equation. Only the zeroth order 
terms in the slow-roll 
approximation will be retained. 

The first two perturbation equations are
\begin{eqnarray}
&&\delta_1\phi^{\prime\prime}-(3r+2)\tau^{-1}\delta_1\phi'
-\partial^2\delta_1\phi+3H^{-1}\tau^{-2}\dot\phi\,\delta_1r
=K\hat\xi,\label{del1}\\
&&\delta_2\phi^{\prime\prime}-(3r+2)\tau^{-1}\delta_2\phi'
-\partial^2\delta_2\phi+3H^{-1}\tau^{-2}\dot\phi\,\delta_2r=\delta_1K\hat\xi\nonumber\\
&&\qquad+3\tau^{-1}\delta_1r\delta_1\phi'
-3ar\tau^{-1}u_1^\alpha\partial_\alpha\delta_1\phi
-3r\dot\phi H^{-1}\tau^{-2}\delta_2\gamma\label{del2}
\end{eqnarray}
To leading order in the slow-roll approximation, the perturbations of $r$ are 
determined entirely by the temperature dependence of the friction coefficient
$\Gamma$. Since $r\propto T^c$ and $\rho_r\propto T^4$, we obtain
\begin{eqnarray}
\delta_1r&=&cr\,{\delta_1\rho_r\over 4\rho_r}\label{dr1}\\
\delta_2r&=&cr\,{\delta_2\rho_r\over 4\rho_r}
-{cr(4-c)\over 2}\left({\delta_1\rho_r\over 4\rho_r}\right)^2.\label{dr2}
\end{eqnarray}
Similarly,
\begin{equation}
\delta K=K{d\ln K\over d\ln T}{\delta_1\rho_r\over 4\rho_r}.
\end{equation}
Before substituting these quantities back into the perturbation equations, it is
convenient to replace perturbations by dimensionless parameters $\zeta_n$ and $\epsilon_n$,
\begin{eqnarray}
\zeta_n&=&{H\,\delta_n\phi\over \dot\phi},\label{zetan}\\
\varepsilon_n&=&{\delta_n\rho_r\over 4\rho_r}.
\end{eqnarray}
On large scales, the parameters $\zeta_1$ and $\varepsilon_1$ become the Bardeen variables 
for an inflaton dominated and a radiation dominated universe respectively. Note that
$\rho_r$ and $\dot\phi$ are related by the slow-roll equation (\ref{sr2}).

After substituting Eqs. (\ref{dr1}) and (\ref{dr2}) into the perturbation 
equations and converting to dimensionless variables we have
\begin{eqnarray}
&&\zeta_1^{\prime\prime}-(3r+2)\tau^{-1}\zeta'
-\partial^2\zeta_1+3cr\tau^{-2}\varepsilon_1
=\hat K\hat\xi,\label{dzeta1}\\
&&\zeta_2^{\prime\prime}-(3r+2)\tau^{-1}\zeta_2'
-\partial^2\zeta_2+3cr\tau^{-2}\varepsilon_2=\delta_1\hat K\hat\xi\nonumber\\
&&\qquad+{\textstyle\frac32}c(4-c)r\tau^{-2}\varepsilon_1^2
+3cr\tau^{-1}\varepsilon_1\zeta'_1
-3ar\tau^{-1}u_1^\alpha\partial_\alpha\zeta_1
-3r\tau^{-2}\delta_2\gamma.
\end{eqnarray}
where $\hat K=HK/\dot\phi$.

\subsection{Radiation fluctuations}

The dominant source of fluctuations in the radiation field 
is an inhomogeneous energy-momentum flux from the inflaton field. 
This transfer of momentum and energy into the radiation is described by
an energy-momentum four-vector $Q_a$, \cite{hwang02,Moss:2007cv}
\begin{equation}
Q_a=-\Gamma u^b\partial_b\phi\,\partial_a\phi,\label{emflux}
\end{equation}
where $u$ is the $4-$velocity of the radiation fluid, $u=(\gamma,u^\alpha)$. 

We shall model the radiation field by a perfect barotropic fluid with 
pressure $p=w\rho_r$, and energy momentum tensor
\begin{equation}
T_{ab}=(1+w)\rho_r u_au_b+p g_{ab},
\end{equation}
and field equations
\begin{equation}
\nabla_aT^{ab}=Q^a.
\end{equation}
The time and space components of the field equations are
\begin{eqnarray}
D\rho_r+(1+w)\rho_r\nabla_au^a&=&Q,\label{fluida}\\
w\partial^\perp_\alpha\rho_r+(1+w)\rho_r Du_\alpha&=&Q^\perp_\alpha\label{fluidb},
\end{eqnarray}
where $\perp$ denotes components perpendicular to $u$ and the source terms are
\begin{eqnarray}
Q&=&-u^aQ_a=\Gamma(D\phi)^2,\label{eflux}\\
Q_\alpha&=&-\Gamma(D\phi)\partial_\alpha\phi.\label{mflux}
\end{eqnarray}

As before, metric perturbations are small on sub-horizon scales and we can use
the cosmological background metric with flat spacial sections. The fluid 
divergence is given by
\begin{equation}
\nabla_au^a=\partial_\alpha u^\alpha+3H\gamma+\dot\gamma
\end{equation} 
Indices are lowered with
the background metric, so that $u_\alpha=a^2u^\alpha$ and $\gamma^2=1+u_\alpha u^\alpha$.

\subsubsection{First order perturbations}

First order perturbations of Eqs. (\ref{fluida}) and (\ref{fluidb}) give
\begin{eqnarray}
\delta_1\dot\rho_r+3H(1+w)\delta_1\rho_r+(1+w)\rho_r\,\partial_\alpha u_1^\alpha
&=&\delta_1Q,\\
w\partial_\alpha\delta_1\rho_r+(1+w)\rho_r\dot u_{1\alpha}
+3H(1+w)\rho_r u_{1\alpha}&=&\delta_1 Q_\alpha.
\end{eqnarray}
Perturbations in the energy and momentum fluxes (\ref{eflux}) and (\ref{mflux}) are caused by
perturbations in the scalar field and perturbations in the friction coefficient, Eq. (\ref{dr1}),
\begin{eqnarray}
\delta_1Q&=&cH\delta_1\rho_r+2\Gamma\dot\phi\,\delta_1\dot\phi\\
\delta_1Q_\alpha&=&-\Gamma\dot\phi\,\partial_\alpha\delta_1\phi.
\end{eqnarray}
From this point on we shall take a radiation fluid with $w=1/3$.

The dimensionless flux perturbation $q_n=\delta_n Q/Q$ can be introduced,  
in addition to the perturbations $\zeta_1$, and $\varepsilon_1$ used previously.
Using the time coordinate $\tau=1/(aH)$,
\begin{eqnarray}
&&a\partial_\alpha u_1^\alpha=3\epsilon'_1-12\tau^{-1}\varepsilon_1+3\tau^{-1}q_1,\\
&&u_1^{\alpha\prime}-5\tau^{-1}u_1^\alpha=a\partial^\alpha\varepsilon_1-3\tau^{-1}q_1^\alpha.
\label{depsilon1}
\end{eqnarray}
where
\begin{eqnarray}
q_1&=&c\varepsilon_1-2\tau\,\zeta_1'\\
q_{1\alpha}&=&-a\tau\partial_\alpha\zeta_1.
\end{eqnarray}
The velocity is given in terms of the inflaton and density fluctuations by
\begin{equation}
a\partial_\alpha u_1^\alpha=3\varepsilon_1'-6\zeta_1'-(12-3c)\tau^{-1}\varepsilon_1
\end{equation}
After eliminating the velocity,
\begin{eqnarray}
&&\varepsilon_1''-(8-c)\tau^{-1}\varepsilon_1'+(20-5c)\tau^{-2}\varepsilon_1
-{\textstyle\frac13}\partial^2\varepsilon_1\nonumber\\
&&\qquad\qquad-2\zeta_1''+8\tau^{-1}\zeta_1'-\partial^2\zeta_1=0.
\end{eqnarray}
This equation was derived previously in Ref. \cite{Graham:2009bf}.

\subsubsection{Second order perturbations}

Second order perturbations of Eqs. (\ref{fluida}) and (\ref{fluidb}) give
\begin{eqnarray}
&&\delta_2\dot\rho_r+3H(1+w)\delta_2\rho_r+(1+w)\rho_r\,\partial_\alpha u_2^\alpha
\nonumber\\
&&\quad+D_1\delta_1\rho_r+(1+w)\delta_1\rho_r\,\partial_\alpha u_1^\alpha
+(1+w)H\rho_r\theta_2=\delta_2 Q\\
&&w\partial_\alpha\delta_2\rho_r+(1+w)\rho_r \dot u_{2\alpha}+3H(1+w)\rho_r u_{2\alpha}
\nonumber\\
&&\quad+w u_{1\alpha}\delta_1\dot\rho_r+(1+w)\delta_1\rho_r\dot u_{1\alpha}
+(1+w)\rho_r D_1u_{1\alpha}=\delta_2Q_\alpha- u_{1\alpha}\delta_1 Q
\end{eqnarray}
where $D_1= u_1^\alpha\,\partial_\alpha$ and 
$H\theta_2=3H\delta_2\gamma+\delta_2\dot\gamma$. 
In terms of dimensionless parameters and $w=1/3$,
\begin{eqnarray}
&&a\partial_\alpha u_2^\alpha=3\varepsilon'_2-12\tau^{-1}\varepsilon_2+3\tau^{-1}q_2\nonumber\\
&&\qquad -4a\varepsilon_1\partial_\alpha u_1^\alpha-3aD_1\varepsilon_1-3\tau^{-1}\delta_2\gamma
+\delta_2\gamma'\\
&&u_2^{\alpha\prime}-5\tau^{-1}u_2^\alpha-a\partial^\alpha\varepsilon_2+3\tau^{-1}q_2^\alpha=\nonumber\\
&&\qquad -u_1^\alpha\varepsilon_1'-4\varepsilon_1 u_1^{\alpha\prime}
+8\tau^{-1}\varepsilon_1 u_1^{\alpha}+aD_1u_1^\alpha
+3\tau^{-1}u_1^\alpha q_1
\end{eqnarray}

The second order momentum flux is
\begin{equation}
q_2^\alpha=-a\tau\partial^\alpha\zeta_2-ca\tau\varepsilon_1\partial^\alpha\zeta_1+
a\tau^2\zeta_1'\partial^\alpha\zeta_1,
\end{equation}
and the second order energy flux is
\begin{equation}
q_2=c\epsilon_2-2\tau\zeta_2'-{\textstyle \frac12}c(4-c)\varepsilon_1^2-2c\tau\varepsilon_1\zeta_1'
+\tau^2\zeta_1^2+2a\tau D_1\zeta_1+2\delta_2\gamma.
\end{equation}
An equation for the energy density fluctuations can be found by eliminating the velocity as before.
We have done this using a computer algebra package, and the result is given in the summary section
below.

\subsection{Metric fluctuations}\label{metricsec}

We have been claiming up to now that the metric fluctuations play no role in the analysis. This can
be checked using the gauge ready equations for metric perturbations from Ref. \cite{hwang02}.
The new quantities we require at first order are the proper time perturbation $\alpha_1$, the scale
factor perturbation $\varphi_1$, the perturbed expansion of the hypersurface normal vectors
$\kappa_1$ and the perturbed shear of the hypersurface normal vectors $\chi_1$. For convenience we
pick a uniform expansion-rate gauge $\kappa_1=0$, although the conclusions hold in for any
reasonable gauge choice.

When the metric perturbations are included and we take the leading order in the slow-roll
approximation, the scalar field equation (\ref{dzeta1}) and the fluid equation (\ref{depsilon1}) 
become
\begin{eqnarray}
&&\zeta_1^{\prime\prime}-(3r+2)\tau^{-1}\zeta'
+k^2\zeta_1+3cr\tau^{-2}\varepsilon_1+\tau^{-1}\alpha_1'-3\tau^{-2}\alpha_1
=\hat K\hat\xi,\label{dphi1}\\
&&u_1^{\alpha\prime}-5\tau^{-1}u_1^\alpha=a\partial^\alpha(\varepsilon_1-\alpha_1)
-3\tau^{-1}q_1^\alpha
\end{eqnarray}
The Raychaudhuri equation gives the metric perturbation $\alpha_1$ 
(see Eq. (14) in \cite{hwang02} or Eq. (A22) in \cite{Moss:2007cv}),
\begin{equation}
\alpha_1={\epsilon\over (1+r)^2}{6r\varepsilon_1-4\tau\zeta_1'+6(1+r)\zeta_1
\over k^2\tau^2-3\epsilon/(1+r)+4\epsilon/(1+r)^2}
\end{equation}
Clearly $\alpha_1<<\varepsilon_1$ and  $\alpha_1<<\zeta_1$, not only for sub-horizon scales
$k\tau>>1$ but also for scales comparable to the horizon when $k\tau\approx 1$. The equations for
the second order metric variation $\alpha_2$ follow a similar pattern, apart from the presence of
a large number of extra quadratic terms in the first order perturbations. The effects of $\alpha_2$
are also suppressed by the slow-roll parameters.

\subsection{Summary}

We now have a full set of perturbation equations up to second order 
which can be used to determine the radiation and scalar field fluctuations
$\varepsilon_n$ and $\zeta_n$. The first order equations were solved numerically 
in Ref \cite{Graham:2009bf}, and an encouraging feature of the numerical work was 
that the results agreed 
well with analytic approximations. We shall be making similar approximations
to the second order equations in the next section.

The first order equations are
\begin{eqnarray}
&&\zeta_1^{\prime\prime}-(3r+2)\tau^{-1}\zeta_1'
-\partial^2\zeta_1+3cr\tau^{-2}\varepsilon_1
=\hat K\hat\xi,\label{dphi1}\\
&&\varepsilon_1''-(8-c)\tau^{-1}\varepsilon_1'+(20-5c)\tau^{-2}\varepsilon_1
-{\textstyle\frac13}\partial^2\varepsilon_1\nonumber\\
&&\qquad\qquad-2\zeta_1''+8\tau^{-1}\zeta_1'-\partial^2\zeta_1=0\label{drho1}
\end{eqnarray}
The fluid velocity is irrotational at this order, and it can be determined by the equation,
\begin{equation}
a\partial_\alpha u_1^\alpha=3\varepsilon_1'-6\zeta_1'-(12-3c)\tau^{-1}\varepsilon_1
\label{veq}
\end{equation}

The second order perturbation equations are
\begin{eqnarray}
&&\zeta_2^{\prime\prime}-(3r+2)\tau^{-1}\zeta_2'
-\partial^2\zeta_2+3cr\tau^{-2}\varepsilon_2=\delta_1\hat K\hat\xi\nonumber\\
&&\qquad+{\textstyle\frac32}c(4-c)r\tau^{-2}\varepsilon_1^2
+3cr\tau^{-1}\varepsilon_1\zeta'_1
-3ar\tau^{-1}u_1^\alpha\partial_\alpha\zeta_1
-3\tau^{-2}r\delta_2\gamma,\label{dphi2}\\
&&\varepsilon_2''-(8-c)\tau^{-1}\varepsilon_2'+(20-5c)\tau^{-2}\varepsilon_2
-{\textstyle\frac13}\partial^2\varepsilon_2\nonumber\\
&&\qquad\qquad-2\zeta_2''+8\tau^{-1}\zeta_2'-\partial^2\zeta_2=
j^{(1)}+j^{(2)}+j^{(3)},
\label{drho2}
\end{eqnarray}
where the source terms $j^{(n)}$ are ordered so that increasing $n$ implies decreasing size for
large $\tau$,
\begin{eqnarray}
j^{(1)}&=&a^2\, \partial^\alpha\left[(c\varepsilon_1-\tau\zeta_1')\partial_\alpha\zeta_1\right]
-{\textstyle\frac13}a^2\partial^\alpha(3\zeta_1+\varepsilon_1)
\partial_\alpha(9\zeta_1+2\varepsilon_1),\label{drhoj}\\
j^{(2)}&=&(2c\varepsilon_1-2\tau\zeta_1')\zeta_1^{\prime\prime}
+27\zeta_1^{\prime 2}-(24-2c)\zeta_1'\varepsilon_1'+6\varepsilon_1^{\prime 2}\nonumber\\
&&+32(3-c)\tau^{-1}\zeta_1'\varepsilon_1-(12-c)(4-c)\tau^{-1}\varepsilon_1'\varepsilon_1\nonumber\\
&&+{\textstyle\frac12}(17c-48)(c-4)\tau^{-2}\varepsilon_1^2+au_1^\alpha\partial_\alpha\omega\\
j^{(3)}&=&-11\tau^{-2}\delta_2\gamma,
\end{eqnarray}
where
\begin{equation}
\omega=-7\zeta_1'+{\textstyle\frac43}\varepsilon_1'+{\textstyle\frac13}(6c-35)\tau^{-1}\varepsilon_1
-11\tau^{-1}\zeta_1.
\end{equation}

\section{Strong dissipation and small scales}\label{secss}

The perturbation equations simplify considerably in the strong regime of
warm inflation. In this regime, the thermal fluctuations are generated on 
small scales, compared to the cosmological horizon, and their amplitude 
approaches a simple power-law behaviour. This contrasts
with vacuum fluctuations, which oscillate on sub-horizon scales and only
freeze-out when they grow larger than the horizon.

Fourrier transforms are defined
with respect to the comoving coordinates $x^\alpha$,
\begin{equation}
\zeta_n(k)=\int d^3x \,\zeta_n(x)e^{i{\bf k}\cdot{\bf x}}
\end{equation}
The fluid velocity decomposes into an irrotational scalar component $u_n$ along ${\bf k}$ and a
solenoidal component $u_n^{T\alpha}$ perpendicular to ${\bf k}$. Only the scalar component, 
\begin{equation}
u_n=ia\hat k_\alpha u_n^\alpha,
\end{equation}
contributes to the scalar density perturbations at second order.
In this section we shall we shall take $r=\Gamma/3H>>1$ and concentrate on physical scales small 
compared to the horizon, where the wave number $k>>aH$.

From Eq. (\ref{drho1}), the first-order density and scalar 
fluctuations are related by
\begin{equation}
\varepsilon_1\approx-3\zeta_1.\label{approx1}
\end{equation}
This approximation can be substituted into Eq. (\ref{del1}). 
\begin{equation}
\zeta_1^{\prime\prime}-(3r+2)\tau^{-1}\zeta_1'
+k^2\zeta_1-9cr\tau^{-2}\zeta_1
=\hat K\hat\xi
\end{equation}
We solve this using a Green function,
\begin{equation}
\zeta_1={1\over k}\int_\tau^\infty G(k\tau,k\tau')(k\tau')^{1-2\nu}\,
\hat K\hat\xi(\tau')\,d\tau',
\label{usol}
\end{equation}
where the retarded Green function $G(z,z')$ is given by
\begin{equation}
G(z,z')={\pi\over 2}z^{\nu} z^{\prime\nu}
\left(J_{\nu+3c}(z)Y_{\nu+3c}(z')-J_{\nu+3c}(z')Y_{\nu+3c}(z)\right)
\quad\hbox{ for }z<z'.\label{gc}
\end{equation}
and $\nu=3r/2$. A useful feature of the solution is that in the range
$1<<k\tau<<\sqrt\nu$ the fluctuation has an approximate power law behaviour,
$\zeta_1\propto \tau^{-3c}$. This allows us to use
\begin{equation}
\zeta_1'\approx -3c\tau^{-1}\zeta_1.\label{approx2}
\end{equation}
We can use both the approximations (\ref{approx1}) and (\ref{approx2}) in the
velocity equation (\ref{veq}), to get
\begin{equation}
k u_1\approx36(1+c)\tau^{-1}\zeta_1.\label{approx3}
\end{equation}

At second order in the perturbation amplitude, the dominant terms for large
$k\tau$ in Eq. (\ref{drho2}) are
\begin{equation}
k^2\varepsilon_2+3k^2\zeta_2=j^{(1)}(k),
\end{equation} 
where $j^{(1)}$ is given by Eq. (\ref{drhoj}). The approximations (\ref{approx1}) and
(\ref{approx2}) imply $j^{(1)}\approx0$ and
\begin{equation}
\varepsilon_2\approx -3\zeta_2.
\end{equation}
We use this together with the approximations (\ref{approx1}), (\ref{approx2}) and (\ref{approx3}) in
Eq. (\ref{dphi2}) to get
\begin{eqnarray}
&&\zeta_2^{\prime\prime}-(3r+2)\tau^{-1}\zeta_2'
+k^2\zeta_2-9cr\tau^{-2}\zeta_2\approx\nonumber\\
&&\qquad A_s\tau^{-2}\zeta_1 *\zeta_1
+A_v\tau^{-2}(k^{-2}k^\alpha\zeta_1)*(k_\alpha\zeta_1)
+A_r\hat K\zeta_1*\hat\xi,\label{delphi2}
\end{eqnarray}
where $*$ denotes a convolution and
\begin{eqnarray}
A_s&=&\frac{27}{2}c(4+c)r,\label{valas}\\
A_v&=&-108(1+c)r,\\
A_r&=&-{3\over 2}(1+c).
\end{eqnarray}
Note that, when the friction coefficient is temperature independent and $c=0$,
then only the $A_v$ and $A_r$ terms survive. These terms agree with the $c=0$ case
investigated in Ref. \cite{Moss:2007cv}, apart form a sign change in $A_v$ caused 
by an error in the sign of the green function used in the earlier work.

The $A_s$ term arises for the temperature dependence of the friction coefficient.
This term is local in space, and we might therefore expect it to produce a local type of
non-gaussianity. The $A_v$ term arises from the fluid advection terms $u^\alpha\partial_\alpha$ in
the perturbation equations, and this term is associated closely with the bulk behaviour of the
radiation field. It distinguishes between vacuum and non-vacuum fluctuations.

\section{Non-gaussianity and the bispectrum}

The small scale inflaton fluctuations develop a simple power-law growth well 
in advance of the time when they cross the horizon. During this stage the
metric fluctuations are relatively small (in the uniform expansion-rate gauge).
Eventually, the wavelength of the perturbations crosses the effective
cosmological horizon and the wavelength grows to a regime where the metric 
perturbations become important. On large scales
it becomes possible to use a small-spatial-gradient expansion (first
formalised by Salopeck and Bond \cite{Salopek:1990jq} and later developed into the `delta-$N$'
formalism \cite{Sasaki:1995aw,Lyth:2004gb,Lyth:2005fi} ). This approach allows
us to define a perturbation $\zeta$ which is is constant on large scales, even
in the full non-linear theory. The bispectrum  and the 
non-linearity of the density fluctuations can be approximated, to a reasonable
accuracy, by matching the small and large scale approximations at horizon
crossing. 

In a uniform spatial curvature gauge, the perturbation $\zeta$ is defined by
\begin{equation}
\zeta=\int_{\phi(\tau)}^{\phi(x,\tau)} {H(\phi')\over\dot\phi(\phi')}d\phi',
\end{equation}
where $H\equiv H(\phi)$ and $\dot\phi\equiv \dot\phi(\phi)$ are given by solving the 
slow-roll equations (\ref{sr1}-\ref{sr3}). When the inflaton perturbations are expanded as
before in eq. (\ref{ipe}), we have 
\begin{equation}
\zeta=\zeta_\phi\delta_1\phi+\zeta_\phi\delta_2\phi+
\frac12\zeta_{\phi\phi}\delta_1\phi*\delta_1\phi+\dots
\end{equation}
where $\phi$ subscripts denote derivatives with respect to $\phi$ and 
$\zeta_\phi=H/\dot\phi$. Note that, according to eqs. (\ref{sr1}-\ref{sr3}), 
$\zeta_{\phi\phi}/\zeta^2_\phi$ can be dropped because it is first order 
in the slow-roll expansion. We therefore have
\begin{equation}
\zeta=\zeta_1+\zeta_2+\dots,
\end{equation}
where the $\zeta_n$ are identical to the parameters defined in Eq. (\ref{zetan}) of the previous
section.

The power spectrum and bispectrum of the density perturbations were defined in the introduction. The
first order perturbations $\zeta_1$ are gaussian fields and
their bispectrum vanishes. The leading order contribution to the bispectrum
must therefore include a contribution to $\zeta$ from the second order 
perturbation,
\begin{equation}
\sum_{\rm cyclic}\langle \zeta_1({\bf k}_1,\tau)
\zeta_1({\bf k}_2,\tau)
\zeta_2({\bf k}_3,\tau)\rangle\approx
(2\pi)^3B_\zeta(k_1,k_2,k_3)
\delta^3({\bf k}_1+{\bf k}_2+{\bf k}_3),\label{bformula}
\end{equation}
where `cyclic' denotes cyclic permutations of 
$\{{\bf k}_1,{\bf k}_2,{\bf k}_3\}$. The second order perturbation can be
obtained by solving Eq. (\ref{delphi2}) using Eq. (\ref{usol}).

We shall split the second order perturbation $\zeta_2$, into three parts
$\zeta_2=\zeta_s+\zeta_v+\zeta_r$, where each part is 
sourced by the corresponding terms with coefficients $A_s$, $A_v$ or $A_r$
in Eq. (\ref{delphi2}). Beginning with the $\zeta_s$ term, we have
\begin{equation}
\zeta_s^{\prime\prime}-(3r+2)\tau^{-1}\zeta_s'
+k^2\zeta_s-9cr\tau^{-2}\zeta_s=A_s\tau^{-2}\zeta_1 *\zeta_1.
\end{equation}
The solution can be obtained using the green function (\ref{gc}),
\begin{equation}
\zeta_s=A_s\int_\tau^\infty k d\tau' G(k\tau,k\tau')(k\tau')^{-1-2\nu}\,
\zeta_1*\zeta_1
\end{equation}
The contribution $B_s$ to the density fluctuation bispectrum from $\zeta_s$ is
\begin{eqnarray}
&&(2\pi)^3B_s(k_1,k_2,k_3)
\delta^3({\bf k}_1+{\bf k}_2+{\bf k}_3)=\nonumber\\
&&\quad A_s
\sum_{\rm cyclic}\int_\tau^\infty k_3 d\tau' G(k_3\tau,k_3\tau')(k_3\tau')^{-1-2\nu}
\langle \zeta_1({\bf k}_1,\tau)
\zeta_1({\bf k}_2,\tau)
\zeta_1*\zeta_1({\bf k}_3,\tau')\rangle
\end{eqnarray}
Now we can use Eq. (\ref{usol}) for $\zeta_1$ and decompose the stochastic average
of the noise terms into products of the correlation function (\ref{cf}). The result is that
\begin{equation}
B_s=2A_s\sum_{\rm cyclic}{\hat K^4\over k_1^3 k_2^3}
\int_\tau^\infty k_3 d\tau' G(k_3\tau,k_3\tau')(k_3\tau')^{-1-2\nu}
F(k_1\tau,k_1\tau')F(k_2\tau,k_2\tau'),
\end{equation}
where
\begin{equation}
F(k\tau_1,k\tau_2)=
k\int_{\tau_2}^\infty d\tau'
G(k\tau_1,k\tau')G(k\tau_2,k\tau')
(k\tau')^{2-4\nu}.
\end{equation}
This integral can be evaluated analytically when $\nu$ is large. The integral is 
expressed in terms the power spectrum $P_\zeta(k,\tau)$ in appendix \ref{appb}, and gives
\begin{equation}
B_s=2A_s\sum_{\rm cyclic}P_\zeta(k_1,\tau)P_\zeta(k_2,\tau)(k_3\tau)^{6c}
\int_\tau^\infty k_3 d\tau' G(k_3\tau,k_3\tau')(k_3\tau')^{-1-2\nu-6c}.
\end{equation} 
This integral can also be found in appendix \ref{appb}, leaving
\begin{equation}
B_s={2A_s\over 9cr}f(c)\sum_{\rm cyclic}P_\zeta(k_1,\tau)P_\zeta(k_2,\tau),
\end{equation}
where we have used $\nu=\Gamma/2H$. The factor $f(c)$ is,
\begin{equation}
f(c)=1-(2\Gamma/H)^{-3c/2}.
\end{equation}
We can usually take $f(c)\approx 1$ unless $c$ is very small. The value of $A_s$ was 
found in the previous section (\ref{valas}), so that
\begin{equation}
B_s=3(4+c)f(c)\sum_{\rm cyclic}P_\zeta(k_1,\tau)P_\zeta(k_2,\tau).
\end{equation}
This part of the bispectrum has a local form, as expected, and the value of the non-linearity
parameter for $B_s$ alone is,
\begin{equation}
f^{local}_{NL}={5\over 2}(4+c)f(c).
\end{equation}
Remarkably, this is only weakly dependent on $\Gamma$. However, some caution
needs to be exercised due to the rather crude nature of the matching procedure applied at the
horizon scale, where the approximations used in Sect. \ref{secss} break down. 

In Ref. \cite{Graham:2009bf}, an alternative approximation scheme was introduced to cover
fluctuations at the horizon scale. This approximation implied that the density fluctuation power
spectra approach constant values at the horizon scale, with $P_\zeta(k_1,\tau)=P_\zeta(k_1)\propto
k^{-3}$ (or $k^{n_s-4}$ at first order in the slow-roll parameters). In the present context, this
suggests that the small scale approximation will extend to the horizon scale when the bispectrum is
expressed in terms of the power spectrum, as we have done above.

\begin{center}
\begin{figure}[htb]
\scalebox{0.4}{\includegraphics{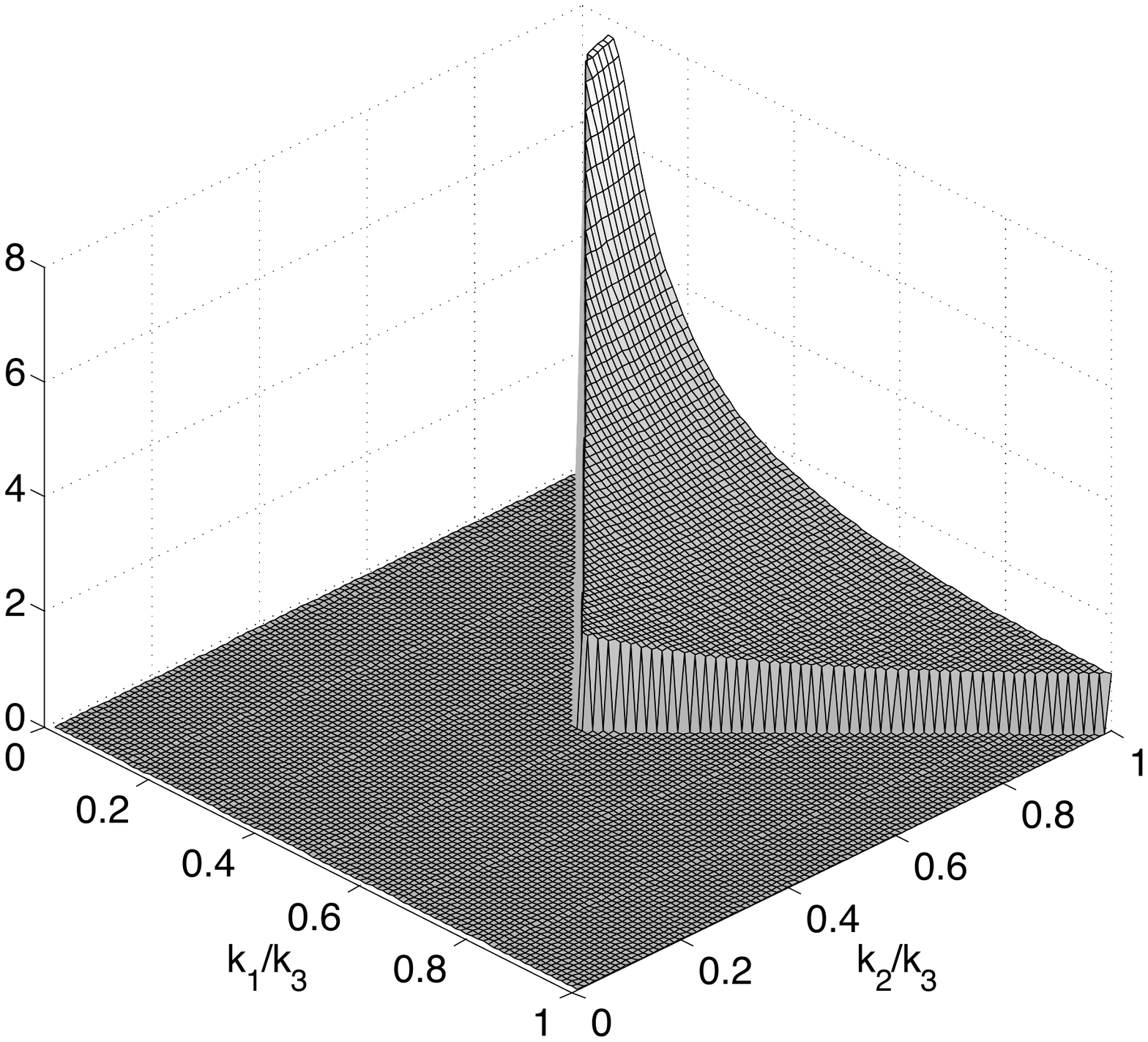}}
\caption{A contour plot of the truncated local bispectrum $B_s$ as a function of $k_1$ and $k_2$. 
The plot shows $B_s(x_1,x_2,1)x_1^2x_2^2\tanh^2(rx_1)\tanh^2(rx_2)/B_s(1,1,1)$, where $x_1=k_1/k_3$,
$x_2=k_2/k_3$ and $r=20$. }
\label{fig0}
\end{figure}
\end{center}

The bispectrum reaches its largest values for oblique wave-vector triangles, with $k_1<<k_2\approx
k_3$ (up to permutations of the sides). However,  if the triangle becomes too oblique, then the
result no longer applies because it is not possible to have $k_1\tau$ and $k_2\tau$ in the same
range where the approximations are valid. The solution \cite{Moss:2007cv} is to cut off the
bispectrum when
$k_2/k_1>r=\Gamma/3H$ (or similarly for any other pairs of sides). There is some residual dependence
of the bispectrum on $\Gamma$ due to this effect. The bispectrum $B_S$ with a cut-off for
$k_2/k_1>r$ is plotted in figure \ref{fig0}.

\begin{center}
\begin{figure}[htb]
\scalebox{0.4}{\includegraphics{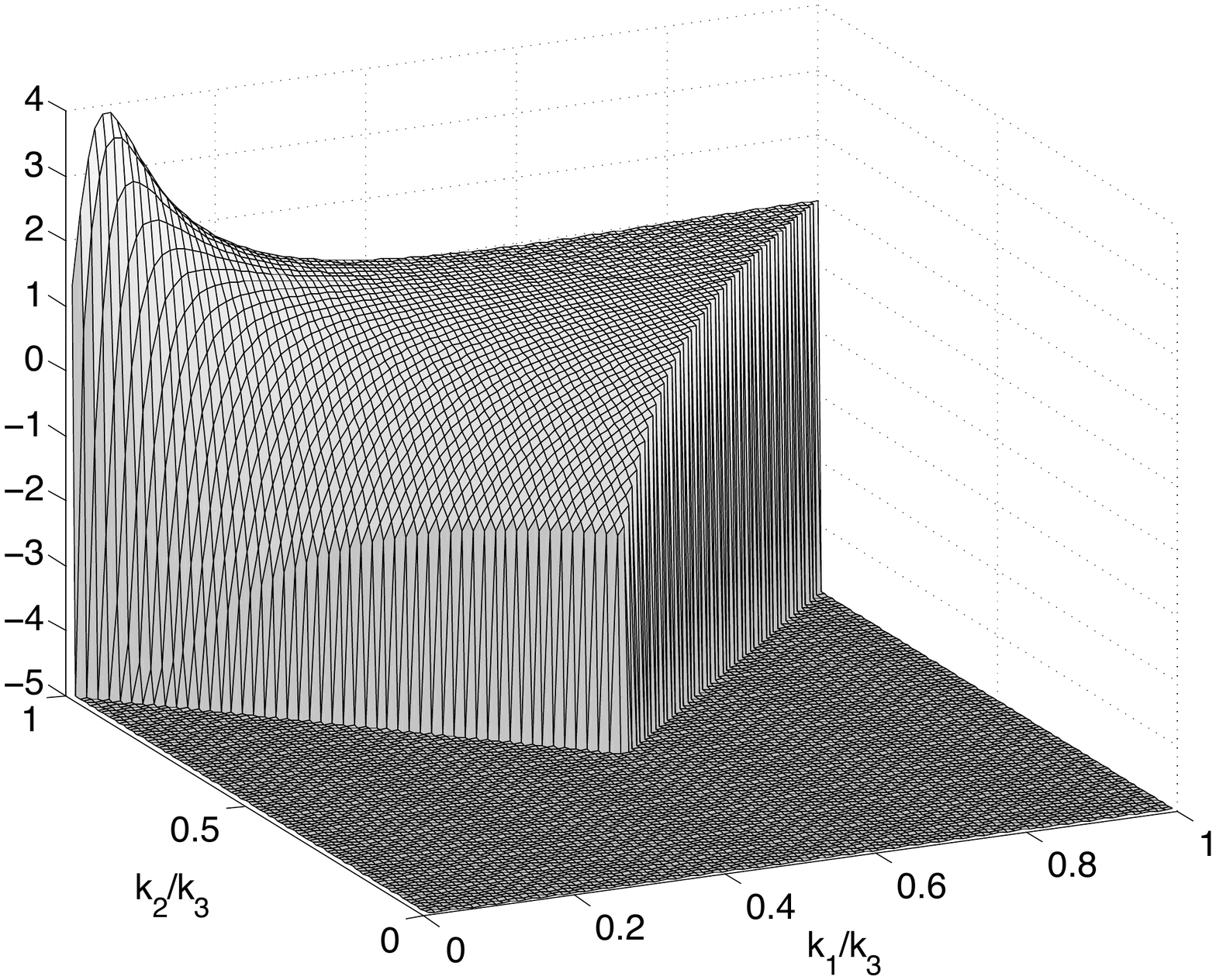}}
\caption{A contour plot of the truncated local bispectrum $B_s$ as a function of $k_1$ and $k_2$. 
The plot shows $B_v(x_1,x_2,1)x_1^2x_2^2\tanh^2(rx_1)\tanh^2(rx_2)/B_v(1,1,1)$, where $x_1=k_1/k_3$,
$x_2=k_2/k_3$ and $r=20$.}
\label{fig1}
\end{figure}
\end{center}

The other contributions to the bispectrum are obtained by following the same steps. The result for
$B_v$ is
\begin{equation}
B_v=-{12\over c}(1+c)f(c)\sum_{\rm cyclic}
(k_1^{-2}+k_2^{-2}){\bf k_1}\cdot{\bf k_2}\,P_\zeta(k_1)P_\zeta(k_2),
\end{equation}
Note that, in the limit $c\to 0$, the function $f(c)\sim (3c/2)\ln(\Gamma/2H)$ and the
result agrees with Ref. \cite{Moss:2007cv}, apart from the overall sign mentioned earlier.
The bispectrum is concentrated on oblique triangle shapes as before, but this time the contribution
to $f_{NL}$ is strongly dependent of the shape of the wave vector triangle and even changes sign as
the triangle becomes more oblique. The dependence of the bispectrum $B_v$ on $k_1$ and $k_2$ for $r=20$
is plotted in Fig. \ref{fig1}. The contribution from the squeezed triangles is reduced relative
to the local case, but the squeezed triangle contribution rises for larger values of $r$

The final contribution to the bispectrum $B_r$ depends on the integral (\ref{fhat}). This
contribution is suppressed by factors of $\Gamma/3H$. It is only relevant when $c\approx 0$, and
even then this contribution is dominated by $B_v$. The total bispectrum is effectively the sum of
two terms,
\begin{equation}
B=\frac65f^{local}_{NL}\sum_{\rm cyclic}P_\zeta(k_1)P_\zeta(k_2)
-\frac65f_{NL}^{adv}\sum_{\rm cyclic}(k_1^{-2}+k_2^{-2})\,{\bf k_1}\cdot{\bf
k_2}\,P_\zeta(k_1)P_\zeta(k_2),
\label{biparas}
\end{equation}
where $f_{NL}^{adv}$ is named after fluid advection terms. The negative sign in front of
$f_{NL}^{adv}$ is chosen so that,  for equilateral triangles, $f_{NL}=f^{local}_{NL}+f_{NL}^{adv}$.

\begin{center}
\begin{figure}[htb]
\scalebox{0.8}{\includegraphics{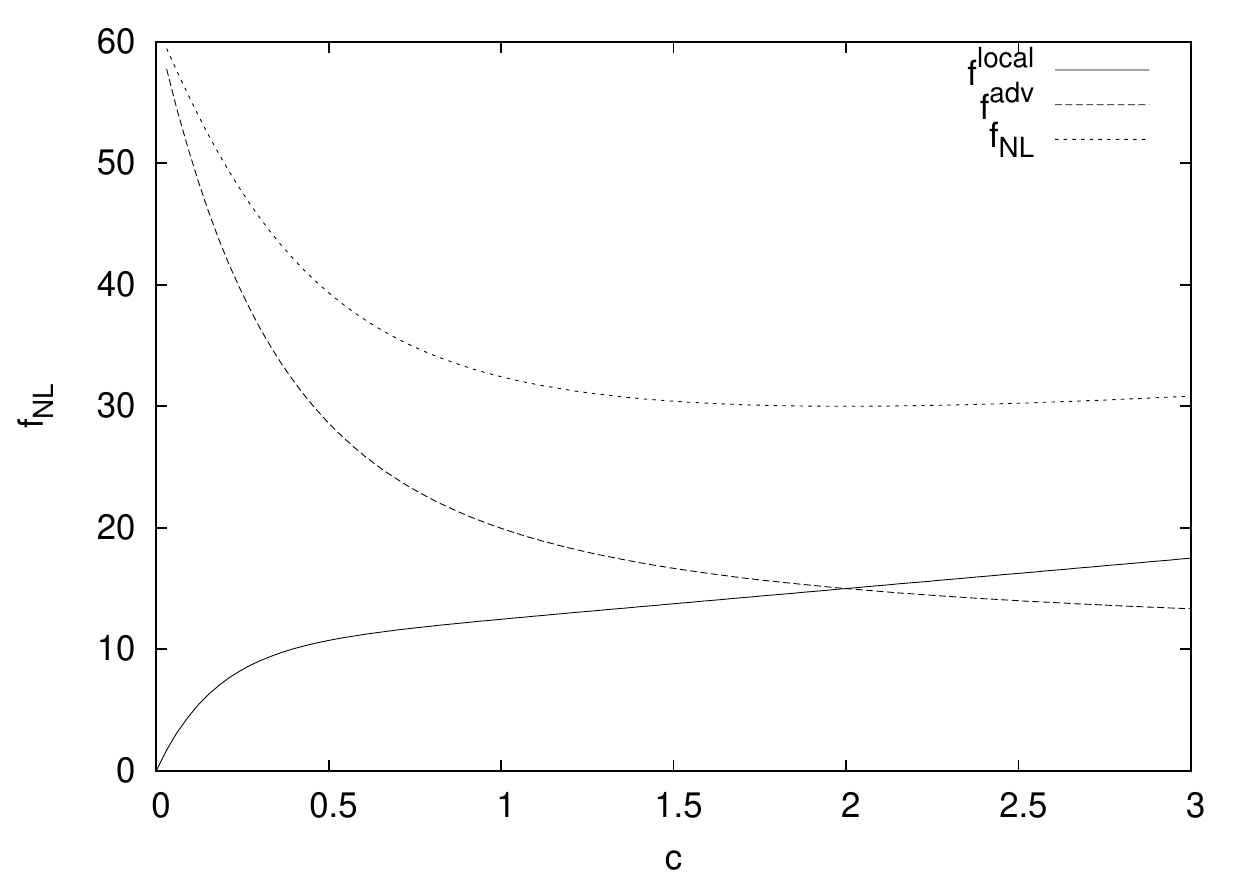}}
\caption{A plot of $f_{NL}$ for equilateral triangles against $c$. The
curves represent two different contributions with different dependence on 
wave numbers. In this plot, $\Gamma/3H=10$ and $\Gamma\propto T^c$.
The contribution from $f^{adv}_{NL}$ should be multiplied by the values in Fig. 1 for other triangle
shapes, and experiments which search for constant (or local) $f^{local}_{NL}$ 
will see predominantly $f_{NL}^{local}$.}
\label{fig2}
\end{figure}
\end{center}

The values of $f_{NL}^{local}$ and $f_{NL}^{adv}$ are plotted in Fig. \ref{fig2}. The contribution
to the bispectrum
from $B_v$ dominates at small values of $c$, but then decreases, whilst the local component $B_s$
becomes larger as $c$ increases. The momentum dependence makes the contribution to $B$ from $B_v$
harder to detect, but since this component is characteristic of warm inflation it would be
desirable to try to isolate this component
in future CMB experiments.

\section{Conclusions and outlook}

The main results of this paper are given in Fig. \ref{fig2}, which shows the amplitudes of two
contributions to the bispectrum for primordial fluctuations which originate in the strong regime of
warm inflation, with friction coefficient $\Gamma\propto T^c$. Contemporary models of warm inflation
typically have $c=3$ \cite{ BasteroGil:2009ec}. The local component with amplitude $f_{NL}^{local}$ 
is easier to detect, and would most likely be seen first. 

It would be very desirable to search for the component of the bispectrum
with amplitude $f^{adv}_{NL}$, since this component is a characteristic feature of fluctuations
which originate from a non-vacuum source. Some idea of the difficulty of measuring both components
of the
bispectrum can be
gauged by
considering an ideal experiment where the only source of noise is the cosmic variance. We can
parameterise the bispectrum as we did in Eq. (\ref{biparas}),
\begin{equation}
B=\sum_i f_i\overline B^i,
\end{equation}
where $i=1,2$ for $f_{NL}^{local}$ of $f_{NL}^{adv}$ respectively. The observations give a set of
spherical harmonic components
$B^{obs}_{l_1l_2l_3}$, with an upper limit $l\le l_{\rm max}$. 
The estimator for the parameter $f_i$ is
\begin{equation}
\hat f_i=\sum_j(F^{-1})_{ij}\sum_{l_1l_2l_3}
{B^{obs}_{l_1l_2l_3}\overline B^j_{l_1l_2l_3}\over 6C_{l_1}C_{l_2}C_{l_3}},
\end{equation}
where $C_l$ are obtained from the power spectrum, and the Fisher matrix
\begin{equation}
F_{ij}=\sum_{l_1l_2l_3}
{\overline B^i_{l_1l_2l_3}\overline B^j_{l_1l_2l_3}\over 6C_{l_1}C_{l_2}C_{l_3}}.
\end{equation}
The Fisher matrix for the relevant models was evaluated in Ref. \cite{Moss:2007qd}, where it was
found that  the standard deviation of the estimator $\hat f_{NL}^{adv}$ is around 5 times larger
than the standard deviation in the estimator $\hat f_{NL}^{local}$. For Planck, the detection limit
for $f^{local}_{NL}$ is expected to be around $5-10$, depending on how successfully the backgrounds
can be removed. This would imply that Planck would only be able to detect the presence of
$f_{NL}^{adv}$ if the
value was at least $25$. The problems arising from cosmic variance could be overcome by taking
higher resolution surveys, but then the issue of backgrounds contributing to $f_{NL}^{adv}$ has to
be addressed. 

There is another regime of warm inflation, the weak regime, where the approximations we have used
here all break down. Nevertheless, the dependence of the bispectrum on the wave-vector triangle
will contain the same components that we found here, with the possibility of two extra components
appearing from terms which we have been able to discard in the strong regime of warm inflation. In
the weak regime we have to resort to numerical investigation of the second order perturbation
equations to determine the bispectrum fully, and this is something we hope to report on at a later
date.

\appendix

\section{Integrals}\label{appb}

We begin with an approximation to the integral
\begin{equation}
I(k\tau)=k\int_\tau^\infty d\tau'\,G(k\tau,k\tau')(k\tau')^{-1-2\nu-6c}
\end{equation}
where $\nu=\Gamma/2H$ and the retarded green function $G$ is given in Eq.
(\ref{gc}). The leading contributions to the integral for large $\nu$ and fixed $\tau$
come from
\begin{equation}
I(k\tau)\approx -{\pi\over 2}(k\tau)^{-3c}(k\tau)^{\nu'}Y_{\nu'}(k\tau)\int_{k\tau}^\infty J_{\nu'}
(z)\,z^{-1-\nu'-3c}dz,
\end{equation}
where $\nu'=\nu+3c$. We also have
\begin{equation}
Y_\nu(z)\sim 
-z^{-\nu}{2^\nu\over\pi}\Gamma_R(\nu),\qquad
J_\nu(z)\sim 
z^{\nu}{2^{-\nu}\over\Gamma_R(\nu+1)}.
\end{equation}
which give
\begin{equation}
I(k\tau)\sim{1\over 6c\nu}(k\tau)^{-6c}.
\end{equation}
The approximations used above fail when $c$ is small. In that case, there is a saddle point
in the integrand which comes to dominate the result. The easiest way to obtain the contribution from
this saddle point is to obtain the result for $c<0$ and extend to $c>0$. When $c<0$,
\begin{equation}
I(k\tau)\approx -{\pi\over 2}(k\tau)^{-3c}(k\tau)^{\nu'}Y_{\nu'}(k\tau)\int_0^\infty J_{\nu'}
(z)\,z^{-1-\nu'-3c}dz.
\end{equation}
The  integral,
\begin{equation}
\int_0^\infty J_{\nu'}(z)\,z^{-1-\nu'-3c}dz=2^{-3c-1-\nu'}
{\Gamma_R(-3c/2)\over\Gamma_R(\nu'+1+3c/2)}.
\end{equation}
For large $\nu$ and small $c$,
\begin{equation}
I(k\tau)\sim -{1\over 6c\nu}(k\tau)^{-3c}(4\nu)^{-3c/2}.
\end{equation}
Combining the two asymptotic results together gives
\begin{equation}
I(k\tau)\sim {1\over 6c\nu}(k\tau)^{-6c}\left(1-(4\nu)^{-3c/2}\right).
\end{equation}

The next integral is
\begin{equation}
F(k\tau)=
k\int_{\tau}^\infty d\tau'
G(k\tau,k\tau')G(k\tau,k\tau').
(k\tau')^{2-4\nu}
\end{equation}
This integral arises when we calculate the power spectrum
using Eq. (\ref{usol}),
\begin{equation}
P_\zeta(k,\tau)=k^{-3}\hat K^2F(k\tau).\label{psf}
\end{equation}
The leading contributions to the integral for large $\nu$ and fixed $\tau$
come from
\begin{equation}
F(k\tau)\approx
{\pi^2\over 4}
(k\tau)^{2\nu}Y_{\nu'}(k\tau)^2
\int_0^\infty J_{\nu'}(z)^2z^{2-2\nu}dz,
\end{equation}
where $\nu'=\nu+3c$. This gives a standard Schafheitlin integral,
\begin{equation}
\int_0^\infty J_{\nu'}(z)^2
z^{2-2\nu}dz=
{1\over2\sqrt{\pi}}{
\Gamma_R(\nu-1)\Gamma_R(3c+3/2)\over\Gamma_R(\nu-1/2)\Gamma_R(2\nu+3c-1/2)}
\end{equation}
where $\Gamma_R$ is the gamma function. 
Hence,
\begin{equation}
F(k\tau)\sim
\sqrt{\pi\over 32\nu}{\Gamma_R(3c+3/2)\over\Gamma_R(3/2)}
\left({2\nu\over k^2\tau^2}\right)^{3c}
\left(1+{k^2\tau^2\over 2\nu}+\dots\right),\label{fapprox}
\end{equation}
for $k\tau<<\nu^{1/2}$. The $O(\nu^{-1})$ terms come from the expansion of $Y_\nu(k\tau)$.

The next integral is
\begin{equation}
F(k\tau_1,k\tau_2)=
k\int_{\tau_2}^\infty d\tau'
G(k\tau_1,k\tau')G(k\tau_2,k\tau')
(k\tau')^{2-4\nu}
\end{equation}
This is evaluated in exactly the same way as $F(k\tau)$,
\begin{equation}
F(k\tau_1,k\tau_2)\sim \left({\tau_1\over\tau_2}\right)^{3c}F(k\tau_1).\label{f2approx}
\end{equation}
We can also use Eq. (\ref{psf}) to express this in terms of the power spectrum,
\begin{equation}
\hat K^2F(k\tau_1,k\tau_2)\sim 
\left({\tau_1\over\tau_2}\right)^{3c}k^3P_\zeta(k,\tau_1)
\end{equation}

Another integral of this type is
\begin{equation}
\hat F(k_1\tau,k_2\tau)=(k_1k_2)^{1/2}\int_{\tau}^\infty d\tau'
G(k_1\tau,k_1\tau')G(k_2\tau,k_2\tau')
(k_1\tau')^{1-2\nu}(k_2\tau')^{1-2\nu}.
\end{equation}
This time, for large $\nu$,
\begin{equation}
\hat F(k_1\tau,k_2\tau)\sim \left({2k_1k_1\over k_1^2+k_2^2}\right)^{(3c+3)/2}
(2\nu)^{-3c/2}F(k_1\tau,k_2\tau).\label{fhat}
\end{equation}
This integral is negligible compared to $F(k_1\tau,k_2\tau)$.

\bibliography{paper.bib,cosper.bib}

\end{document}